\documentstyle[preprint,aps]{revtex}
\tightenlines
\begin{document}

\newcommand{\beq}{\begin{equation}}
\newcommand{\eeq}{\end{equation}}
\newcommand{\bea}{\begin{eqnarray}}
\newcommand{\eea}{\end{eqnarray}}
\newcommand{\ba}{\begin{array}}
\newcommand{\ea}{\end{array}}
\newcommand{\om}{(\omega )}
\newcommand{\bef}{\begin{figure}}
\newcommand{\eef}{\end{figure}}
\newcommand{\leg}[1]{\caption{\protect\rm{\protect\footnotesize{#1}}}} 

\newcommand{\ew}[1]{\langle{#1}\rangle}
\newcommand{\be}[1]{\mid\!{#1}\!\mid}
\newcommand{\no}{\nonumber}
\newcommand{\etal}{{\em et~al }}
\newcommand{\geff}{g_{\mbox{\it{\scriptsize{eff}}}}} 
\newcommand{\da}[1]{{#1}^\dagger}
\newcommand{\cf}{{\it cf.\/}\ }
\newcommand{\ie}{{\it i.e.\/}\ }
\newcommand{\eg}{{\it e.g.\/}\ }

\title{Quantum teleportation criteria for continuous variables}
\author{Philippe Grangier and Fr\'ed\'eric Grosshans}
\address{Laboratoire Charles Fabry, Institut d'Optique Th\'eorique et Appliqu\'ee, \\
F-91403 Orsay, France}

\maketitle

\begin{abstract}

We discuss the criteria presently used for evaluating the efficiency of quantum teleportation
schemes for continuous variables.
It is argued that the fidelity criterion used so far has some severe drawbacks,
and that a fidelity value larger than 2/3 is actually required for successful 
quantum teleportation. This value has never been reached experimentally so far.

\end{abstract}

%%%%%%%%%%%%%%%%%%%%%%%%%%%%%%%%%%%%%%%%%%%%%%%%%%%%%%%%%%%%%%%%%%%%%%%%%%%%%%%%%%%%%%%%%%%%%%%%%%%%%%%%%%%%
%%%%%%%%%%%%%%%%%%%
\section {Introduction}

Quantum teleportation has emerged in recent years as a major paradigm of 
theoretical \cite{BBC93} and experimental \cite{BPM,BBM,FLB98} quantum information. 
The initial approaches using discrete variables \cite{BBC93,BPM,BBM} have been extended
to continuous quantum variables \cite{FLB98,BK98,RL98}. Though there is a general agreement
about the main ideas of quantum teleportation, some discussions have appeared about
practical details, often related to practical considerations :
``conditional" vs ``unconditional" teleportation \cite{BK98b}, or various - and somehow different - 
sets of criteria for evaluating the efficiency of realistic -and thus
imperfect - teleportation experiments \cite{BK98,RL98}. In this paper, we will review 
these teleportation criteria for continuous quantum variables. We will relate 
these criteria to the various ones introduced in the literature,
and also to the ones previously introduced for characterizing non-ideal 
quantum non-demolition (QND) measurements \cite{HCW90,PRG94,GLP}. Our main conclusion will be 
that the fidelity criterion, which has been the most widely used so far,
is subject to some ambiguities. If particular, we will argue that the quantum vs
``classical" limit $F_{quant} > 0.5$ is at least partially unwarranted, while taking
$F_{quant} > 2/3$ would be much safer - even if more difficult to reach.

Shared entanglement is known to be the basic ingredient for quantum teleportation.
The main line of our argument will be to look in detail at what is actually meant
by ``entanglement", especially in the case where what is shared between
the communicating parties (Alice and Bob) are mixed states rather than pure states. 
There are various ways to define entanglement, depending whether one insists
on the mathematical structure of the state - we will call that aspect ``entanglement"
and use the standard tools of quantum information theory - or on the 
physical properties associated with non-locality and the violation of Bell's inequalities (BI) 
- we will call that aspect ``non-separability". 
As long as Alice and Bob share pure states, the situation is relatively clear  : 
it is known that entanglement and BI violation are equivalent properties of a 
pure quantum state. We point out however that the violation is not necessarily obtained
on the same variables as the teleported ones. For instance, there is no
violation of BI for the continuous variables teleported using an EPR state \cite{FLB98}; 
the violation shows 
up on other variables, such as the parity of the photon number, 
which are much more difficult to access. We will come back to that point later,
but one may consider this as a relatively minor difficulty : the shared 
entangled state does exhibit in principle both entanglement and BI violation.

In practice, due to imperfect transmissions, shared states are most often mixed states. 
Then the above nice equivalence disappears, and it becomes possible
to violate the ``classical" boundary of teleportation, without any violation of BI
\cite{Pr}. This means also that the quantum state - even if it is formally
``entangled" - can be mimicked by a local hidden variable model.
One may then argue that an essential feature of entanglement is lost : though
the quantum entanglement can still not be created locally,
it can be simulated by a {\it local} classical process. 
One may thus seriously question whether a state used is such conditions
would be fully secure, \eg in a quantum cryptography protocol: 
the laws of physics will not forbid a smart eavesdropper to set up a ``fake" transmission line,
using local hidden variables that she fully controls.

A significant extra difficulty with continuous variables schemes is that,
as said above, there is no direct practical way to check whether or not BI are violated.
There is however a weaker test of non-separability, which is to use the 
Einstein-Podolsky-Rosen (EPR) argument itself \cite{epr}. This argument applies whenever
two (non-commuting) measurements on a system allows one to deduce the values
of two other (non-commuting) observables of another remote system, in such
a way that the product of the two inferred variances apparently violate Heisenberg 
inequalities (HI). This violation is apparent only, because the inferred variances 
are actually ``conditional variances", with the condition - \ie the measurement on the 
first system - being changed between the two terms of the product.
Therefore, in the absence of efficient BI test,
we will propose to use this apparent violation of HI as a more effective - and in some 
sense ``necessary" - way to characterize 
the shared entanglement. We will show below that this condition can be recasted
as a condition about the teleportation efficiency. When the two conditional variances
are equal, the condition expressed as a fonction of the usual fidelity
criterion is $F>2/3$. It has never been reached experimentally so far.

\section{Quantum measurement and re-creation schemes}

We assume that joint measurements of both quadratures are performed
on a light beam. We will carry out a linearized, gaussian noise
analysis, which is relevant for the present experimental schemes.
Such an approach has already been used extensively for characterizing
optical QND measurements \cite{PRG94,GLP}.
Denoting as $  X_s^{in}, \;   Y_s^{in}$ the input quadratures,
$  M_X, \;   M_Y$ the measurement results, one has in the general case :
\bea
  M_X &=& g_X   X_s^{in} + f_X   Y_s^{in} +  B_X \nonumber \\
  M_Y &=& f_Y   Y_s^{in} + g_Y   Y_s^{in} +  B_Y 
\eea
where the $f's$ and $g's$ are linearized gain coefficients, and the $B's$
are added noises during the measurement process (with zero mean values).
For simplicity, we will assume that the measurement is good enough to give information of one 
quadrature only, \ie $f_X = f_Y=0$ (this can be obtained from an appropriate quadrature rotation).
One has thus : 
\bea
  M_X &=& g_X   X_s^{in}  + B_X \nonumber \\
  M_Y &=& g_Y   Y_s^{in}  + B_Y 
\eea
Making joint measurements means that $M_X$ and $M_Y$ are compatible, 
and thus commuting quantities. Taking into account that the measurement noise 
is unrelated with the signal input, one has :
\bea
[M_X, \; M_Y] = 0 &=& g_X g_Y [  X_s^{in}, \;    Y_s^{in}]  + [B_X, B_Y]  \nonumber \\
&=& 2 i g_X g_Y + [B_X, B_Y]
\eea
One obtains thus the Heisenberg relation :
\beq
\Delta  B_X   \Delta B_Y \geq | g_X g_Y |
\eeq
It is convenient to define the variances of the 
equivalent input noises \cite{PRG94} associated with the measurements :
\bea
N_X^{m} = (\Delta M_X^{in}/|g_X|)^2  - (\Delta X_s^{in})^2 = (\Delta B_X/|g_X|)^2 \nonumber \\
N_Y^{m} = (\Delta M_Y^{in}/|g_Y|)^2  - (\Delta Y_s^{in})^2 = (\Delta B_Y/|g_Y|)^2 
\eea
One has thus : 
\beq
N_X^{m} N_Y^{m} \geq 1
\label{heisx}
\eeq
This Heisenberg-type relation sets a limit to the simultaneous measurement of 
both quadratures. It can be rewritten in another way by using the input-output
transfer coefficient $T_i^m = R_{i}^m / R_{i}^{in}$
where $i=X, \; Y$, and the $R's$ are the signal to noise ratios of the corresponding channels. 
The transfer coefficient $T$ expresses how well the signal to noise ratio $R$ is processed from 
the input to the output; $T$ is 1 for a perfect system, and 0 if the information is 
totally destroyed. It can be shown simply \cite{PRG94} that :
\beq
T_X^m = \frac{(\Delta X_s^{in})^2}{(\Delta X_s^{in})^2 + N_X^{m}} \; \; \; \;
T_Y^m = \frac{(\Delta Y_s^{in})^2}{(\Delta Y_s^{in})^2 + N_Y^{m}}
\eeq
Assuming that the input signal is in a minimum uncertainty state 
(\ie $\Delta  X_s^{in} \Delta  Y_s^{in} =1$), it can be shown \cite{RL98} that eq. \ref{heisx}
is equivalent to :
\beq
T_X^m + T_Y^m \leq 1
\eeq
The total information transfer in the joint measurement of both quadratures 
has thus an upper bound, which is just another way to write the Heisenberg inequality 
given by eq. \ref{heisx}. 

In a teleportation scheme, the non-perfect classical information which is obtained from the previous
measurement is used to reconstruct the initial state. One has thus :
\bea
  X_s^{out} &=& h_X   M_X +  C_X \nonumber \\
  Y_s^{out} &=& h_Y   M_Y +  C_Y 
\eea
where th $h's$ and $C's$ are again gains and noises. 
Using the same reasoning as above, one obtains immediately : 
\beq
\Delta C_X  \Delta C_Y \geq 1
\eeq
On the other hand, one has : 
\bea
  X_s^{out} &=& h_X g_X   X_s^{in}  + h_X B_X +  C_X =  h_X g_X   X_s^{in}  + D_X\nonumber \\
  Y_s^{out} &=& h_Y g_Y   Y_s^{in}  + h_Y B_Y +  C_Y =  h_Y g_Y   Y_s^{in}  + D_Y 
\label{telinout}
\eea
and thus : 
\beq
 \Delta D_X  \Delta D_Y \geq |1 - h_X g_X h_Y g_Y|
\label{heisd}
\eeq
If the rhs of eq. \ref{heisd} is zero, then there is no lower bound on the product of the
added noise, and perfect reconstruction is possible : this is 
the principle of quantum teleportation. In practice,  the detection and reconstruction 
scheme include a classical part which can be subjected to arbitrary 
gain. It is thus always possible to adjust the parameters in such a way that :
\beq
g_T = h_X g_X = h_Y g_Y = 1
\eeq
which yields a zero lower bound in eq. \ref{heisd}. 
Again like above, one can introduce :
\beq
T_i^{out} = \frac{R_{i}^{out}}{R_{i}^{in}} 
\eeq
Since the added noise may be zero, it is possible to reach $T_X^{out} + T_Y^{out} = 2$
by using teleportation. On the other hand, by using only the available classical information, 
one would be bounded to $(T_X^{out} + T_Y^{out})_{class} = T_X^m + T_Y^m \leq 1$.

Another way to look at this bound is by introducing equivalent input noises. In the 
interesting case where $h_X g_X = h_Y g_Y = 1$, one has simply $N_X^{out} = (\Delta  D_X)^2$ and
$N_Y^{out} = (\Delta  D_Y)^2$.
In order to obtain $T_X^{out} + T_Y^{out} > 1$, one needs actually 
to have $N_X^{out} N_Y^{out}  < 1$. This is certainly possible, since $D_X$ and $D_Y$ are
not the quadrature components of a same mode, but this requires
entanglement, which is the basic ingredient of quantum teleportation. 
We will show below that this condition is equivalent to the apparent HI
violation quoted above, and we will take it as our main teleportation criterion.

As an example, we quote here again the principle of the EPR
teleportation of ref. \cite{FLB98}. By taking again  $h_X g_X = h_Y g_Y = 1$,
and denoting $B'_X = B_X/g_X, \; B'_Y = - B_Y/g_Y$, one has : 
\bea
  X_s^{out} &=&   X_s^{in}  + B'_X +  C_X \nonumber \\
  Y_s^{out} &=&   Y_s^{in}  - B'_Y +  C_Y 
\label{epr}
\eea
One may thus choose $B'_X = X_1$, $C_X=X_2$, $B'_Y = Y_1$, $C_Y=Y_2$, where $1$ and $2$
denote two EPR beams generated in such a way that $(X_1 + X_2)$ and $(Y_1 - Y_2)$ 
are both squeezed \cite{FLB98} ($(Y_1 + Y_2)$ and $(X_1 - X_2)$ are then anti-squeezed).
This warrants the efficiency of the scheme, where $N_X^{out}   N_Y^{out}$
can be arbitrarily low, restricted by practical considerations such as the available squeezing
and the quantum efficiency of the transmission channel (see also discussion in section \ref{dd}). 

As a conclusion, a simple criterion for efficient quantum teleportation is simply that
$T_X^{out} + T_Y^{out} > 1$, {\it provided that the gain for both quadratures is set to 1}.
This can equivalently be written as an Heisenberg product relative to the equivalent input
noises $N_X^{out} N_Y^{out} < 1$. 

\section{Relations to other criteria}
\subsection{Ralph and Lam criteria}
The above criteria are very close from the ones initially proposed by Ralph and Lam \cite{RL98},
since all of them are derived from the criteria used previously 
for characterizing QND measurements \cite{PRG94}. The $T_X^{out} + T_Y^{out} > 1$ criterion 
was actually introduced in ref. \cite{RL98}, where it was also pointed out 
that it is possible to ``cheat" this criteria is the signal gain is much larger than one. 
It is thus crucial to impose a limitation on the gain,
but instead of the unity gain condition used above,
Ralph and Lam consider an input-output conditional variance. In our opinion, this
causes several problems. First, an input-output correlation cannot be simply measured, because
it requires to know the quantum fluctuations of the input beam. Even if these fluctuations are
measured, using \eg a perfect QND measurement, the input beam will be strongly altered,
and the properties of the teleportation may be completely changed. This is actually for avoiding such
problems that the transfer coefficients were introduced for QND measurements (as far
as the conditional variance is concerned, the QND criteria use an output-output 
conditional variance, which is easy to measure \cite{PRG94}). 
Since the direct measurement of the input-output conditional variance is not feasible,
a possible solution is to evaluate it from the transfer coefficients \cite{RL98}.
However, the two criteria will then give a partly redundant information,
in a way which not easy to disintricate. 

A more serious problem is that the criteria of ref. \cite{RL98} may lead to the conclusion
that teleportation can work better when the signal gain significantly deviates from one. 
This is in contradiction with the fidelity criterion, which is 
strongly peaked with a maximum for unity gain (see below). A unity gain actually
insures that the {\it classical} part of the signal will be teleported correctly. 
It is very easy to check whether or not it is satisfied, simply by teleporting
coherent states with various complex amplitudes : for this simple test to be passed,
it is crucial to impose the unity gain condition. We note that
this condition is actually easy to fulfill, since
it is essentially an adjustment in the ``classical" part of the set-up,
\ie turning an electronics circuit knob. 
There is thus no reason to deviate from that condition, and this is why
we take the value of $T_X^{out} + T_Y^{out}$
for unity gain (\ie $g_T = h_X g_X = h_Y g_Y = 1$) as the actual criterion of interest \cite{rem}.

\subsection{Fidelity}
A simple approach to the fidelity criterion can be given when one tries to teleport
an arbitrary coherent state $| \alpha \rangle$. The density matrix of the teleported state
can be expanded on a coherent state basis $| \beta \rangle$, where the probability
to reconstruct the state $| \beta \rangle$ is denoted as $P(\beta)$.

The fidelity is then simply :
\bea
F &=& \int{d^2 \beta P(\beta) | \langle \beta | \alpha \rangle |^2} \no \\
&=& \int{dx \; dy  P(x,y) \exp \left( -\frac{(x-x_a)^2}{4} - \frac{(y-y_a)^2}{4} \right)}
\eea
where $\alpha = (x_a + i y_a)/2$, $\beta = (x + i y)/2$, and the vacuum noise variance
has been normalized to 1. In a gaussian noise hypothesis,
one has :
\bea
P(x,y) &=& \frac{1}{2 \pi \sqrt{N_X^{out}  N_Y^{out}}}
\exp \left( -\frac{(x-x_b)^2}{2 N_X^{out}} - \frac{(y-y_b)^2}{2 N_Y^{out}} \right)
\eea
where the variances of the $x$ and $y$ distributions
are just the equivalent input noise calculated above, and
$(x_b, \; y_b)$ are the mean coordinates of the reconstructed distribution. 
By carrying out the integration one obtains : 
\beq
F  = \frac{2}{\sqrt{(2+N_X^{out})(2+N_Y^{out})}} \exp \left( -\frac{(x_a-x_b)^2}{2(2+N_X^{out})} - 
\frac{(y_a-y_b)^2}{2(2+N_Y^{out})} \right)
\eeq
The fidelity is thus strongly peaked on the condition $(x_a = x_b, \; y_a = y_b)$, 
which is obtained for unity gain ($g_T=1$)in the teleportation scheme. One gets thus \cite{FLB98,BK98} :
\beq
F_{g_T=1} = \frac{2}{\sqrt{(2+N_X^{out})(2+N_Y^{out})}}
\eeq
This quantity is clearly relevant for characterizing quantum teleportation,
and can reach the value $F_{g_T=1} = 1$ when $N_X^{out} = N_Y^{out} =0$, \ie 
when the teleportation noise is zero. A more subtle question is to decide
about a boundary for successful teleportation.
According to ref. \cite{FLB98}, this boundary should be $F_{g_T=1} = 0.5$, which
is obtained for instance when $N_X^{out}=N_Y^{out}=2$, \ie when the two noises associated
with the measurement and reconstruction independantly reach the shot-noise limit. As we have said above, 
we will show now in more details that a more meaningful limit is to take
$N_X^{out} N_Y^{out}=1$, or $N_X^{out}=N_Y^{out}=1$ for equal noises, giving 
a more stringent quantum limit $F_{g_T=1} > 2/3$.

\subsection{Heisenberg criterion}
We give here the mathematical formulation of the main point of this paper,
which is that the EPR argument can be used as a teleportation criterion. 
For this calculation, it is convenient to rewrite eq. \ref{telinout} under the following form,
where unity gain is assumed : 
\bea
  X_s^{out} &=&   X_s^{in}  + X_m + X_r \nonumber \\
  Y_s^{out} &=&   Y_s^{in}  + Y_m + Y_r
\eea
where the subscripts $m$ and $r$ denote respectively the added noises due to
the measurement and the reconstruction. 
We denote $v_A = (\Delta  A)^2$ the variance for each operator, and 
$c_{A,B} = \langle   A   B \rangle$  the correlation between $A$ and $B$,
which is taken real.
As a criteria for non-separability, we will use the EPR argument : 
two different measurements prepare two different states,
in such a way that the product of conditional variances
(with different conditions) violates the Heisenberg principle. 
The relevant conditional variances can be written \cite{PRG94} :
\bea
V_{Xr|Xm} &=& (v_{Xr}  - c_{Xm,Xr}^2/v_{Xm}) \no \\
V_{Yr|Ym} &=& (v_{Yr}  - c_{Ym,Yr}^2/v_{Ym}) \no \\
V_{Xm|Xr} &=& (v_{Xm}  - c_{Xm,Xr}^2/v_{Xr}) \no \\
V_{Ym|Yr} &=& (v_{Ym}  - c_{Ym,Yr}^2/v_{Yr})
\eea
and the classical limit of no apparent violation of HI is :
\beq
V_{Xr|Xm} V_{Yr|Ym} \geq 1 \; \; \; \; \;
V_{Xm|Xr} V_{Ym|Yr} \geq 1
\eeq
Defining $v_{Cx} = (v_{Xr} v_{Xm} - c_{Xm,Xr}^2)$ and $v_{Cy} = (v_{Yr} v_{Ym} - c_{Ym,Yr}^2)$
these two inequalities become :
\beq
v_{Cx} v_{Cy}  \geq v_{Xm} v_{Ym} \; \; \; \; \;
v_{Cx} v_{Cy}  \geq v_{Xr} v_{Yr}
\eeq
It will be useful for the following to note that the lhs of these inequalities 
can be written: 
\bea
&(&v_{Xr} v_{Xm} - c_{Xm,Xr}^2) (v_{Yr} v_{Ym} - c_{Ym,Yr}^2) = \no \\
&(&(v_{Xr} + v_{Xm})^2 - (v_{Xr} - v_{Xm})^2 - 4 c_{Xm,Xr}^2) ((v_{Yr} + v_{Ym})^2 - (v_{Yr} - v_{Ym})^2 - 4 c_{Ym,Yr}^2)/16 = \no \\
&(&v_{Xr} + v_{Xm} +2 c_{Xm,Xr}) (v_{Xr} + v_{Xm} -2 c_{Xm,Xr}) \times \no \\
&(&v_{Yr} + v_{Ym} + 2 c_{Ym,Yr}) (v_{Yr} + v_{Ym} -2 c_{Ym,Yr})/16  \no \\
&-& (v_{Xr} - v_{Xm})^2 v_{Cy}/4 -  (v_{Yr} - v_{Ym})^2 v_{Cx}/4 - (v_{Xr} - v_{Xm})^2 (v_{Yr} - v_{Ym})^2/16 
\eea
The product of the total equivalent input noises is :
\beq
N_X^{out} N_Y^{out} = (v_{Xm} + v_{Xr} + 2 c_{Xm,Xr})(v_{Ym} + v_{Yr} + 2 c_{Ym,Yr}) 
\eeq
Due to the positive character of the variances, one has :
\bea
(v_{Xr} + v_{Xm} - 2 c_{Xm,Xr}) &\leq& 2(v_{Xr} + v_{Xm}) \no \\
(v_{Ym} + v_{Yr} - 2 c_{Ym,Yr}) &\leq& 2(v_{Ym} + v_{Yr}) 
\eea
Combining all the above equations, one obtains :
\bea
N_X^{out} N_Y^{out} &\geq& \frac{2 (v_{Xm} v_{Ym} + v_{Xr} v_{Yr}) + (v_{Xr} - v_{Xm})^2 v_{Cy} + 
(v_{Yr} - v_{Ym})^2 v_{Cx} }{(v_{Xr} + v_{Xm})(v_{Ym} + v_{Yr})} \no \\
&\geq& 1 +  \frac{n}{(v_{Xr} + v_{Xm})(v_{Ym} + v_{Yr})}
\eea 
where $n = (v_{Xm} - v_{Xr})( v_{Ym} - v_{Yr}) + (v_{Xr} - v_{Xm})^2 v_{Cy} + (v_{Yr} - v_{Ym})^2 /v_{Cy}$.
The minimum of $n$ with respect to $v_{Cy}$ is obtained for $v_{Cy} = |(v_{Yr} - v_{Ym})/(v_{Xr} - v_{Xm})|$
and takes the value
\beq 
n = (v_{Xm} - v_{Xr})( v_{Ym} - v_{Yr}) + 2 |(v_{Xm} - v_{Xr})( v_{Ym} - v_{Yr})|
\eeq 
The value of $n$ is thus always positive (or zero), and  one obtains finally : 
\beq
N_X^{out} N_Y^{out} \geq 1
\eeq
as the condition for no useful entanglement between the two beams. 
As shown above, this condition has the nice advantage of being equivalent
to the transfer criterion $ T_X^{out} + T_Y^{out} \leq 1$ when the teleportation input is a minimum
uncertainty state.
For symmetrical noise variance $N_X^{out} = N_Y^{out} \geq 1$, 
it corresponds to the fidelity value : 
\beq
F = 4/ \sqrt{\left(2 + N_X^{out} \right) \left(2 + N_Y^{out} \right)} \leq 2/3
\eeq

\section{Discussion}
\label{dd}
\subsection{An example : sharing imperfect EPR beams}
In order to illustrate the physical significance of these various criteria, we consider
again teleportation using an EPR state \cite{FLB98}, obtained
by recombining two squeezed beams with equal variances $V_{sq1}=V_{sq2}=s < 1$.
The resulting EPR  state is distributed to Alice and Bob
with a quantum efficiency $\eta$, which is supposed to be the same
on the measurement and reconstruction channels (this corresponds to the most favourable
hypothesis). 
This situation provides a simple model of a mixed state where our analysis is relevant,
and it also corresponds to practical experimental situations.

Using the scheme introduced in eq. \ref{epr}, one obtains :
\bea
  X_s^{out} &=&   X_s^{in}  + \sqrt{2 \eta} \; X_{sq1} + \sqrt{1-\eta} \; X'_v +  \sqrt{1-\eta} \; X_v \nonumber \\
  Y_s^{out} &=&   Y_s^{in}  + \sqrt{2 \eta} \; Y_{sq2} - \sqrt{1-\eta} \; Y'_v +  \sqrt{1-\eta} \; Y_v
\eea
where the subscript $v$ denotes vacuum modes. One has thus :
\bea
N_X^{out} &=& N_Y^{out} = 2 (1-\eta + \eta s) \no \\
T_X^{out} + T_Y^{out} &=&  2/(3 -2 \eta +2 \eta s) \no \\
F &=& 1/(2 -\eta + \eta s)
\eea
These various curves are plotted on Fig. 1. 
For s=0 (perfect squeezing), the product 
$N_X^{out} N_Y^{out}$ goes below 1 for $\eta > 1/2$, for which value
$T_X^{out} + T_Y^{out} = 1$ and $F=2/3$. 
According to our criteria, successful teleportation requires both that
the transmission efficiency $\eta$ is larger than 50 \%, 
and that the noise reduction is larger than 3 dB.
On the other hand, the ``classical" limit $F=0.5$
is beaten as soon as the efficiency and squeezing are not vanishingly small.
This clearly shows again that the $F=0.5$ limit is very loose,
and appears as a criteria for the use of squeezed light, rather than for the existence of 
quantum non-separability.

\subsection{Some more remarks}

It is worth noticing that fidelity and efficiency are usually 
considered as two different concepts :
if one is able to ``distillate" entangled pure states from shared mixed states,
then higher fidelity can be obtained at the expense of a lower efficiency. 
However, this idea does not correspond to the experimental situation
described in ref. \cite{FLB98}, where the imperfect EPR state is directly used
for the teleportation. In other terms, we claim that as long as the 
shared state used for the reconstruction has not been purified to an entangled pure state,
any ``classical" limit rests on a shaky ground, unless it is completed
by a supplementary EPR-type or Bell-type argument as we have done here.

The situation for pure states can be examined by taking $\eta=1$ in the simple model above.
The shared state is then a pure EPR state, and $F>0.5$ as soon as some squeezing
is available. On the other hand, our criteria still requires more that 3 dB squeezing ($s < 0.5$).
The physical meaning of this condition is clear : the individual EPR beams are very noisy,
and $s < 0.5$ is required to bring the conditional variances in the HI-violating domain \cite{epr}.
On the other hand, if $s \geq 0.5$, though the state is in principle (non-locally) entangled,
the behaviour of the {\it observed} quantities can be mimicked by a {\it classical}
and {\it local} model, which does not involve non-separability in the EPR sense.
Whether or not this is acceptable for ``successful teleportation" is an open question;
our present answer is clearly no, but the relationship between teleportation and non-separability
is still a subject of active discussions (see \eg \cite{pp} and references therein). 

\section{Conclusion}

The main goal of this paper is to define a teleportation criteria for
continuous variables, based upon the notion of non-separability, defined here
as the violation of Heisenberg inequalities for products of conditional variances.
The obtained condition reads $N_X^{out} N_Y^{out} < 1$, 
where $N_X^{out}$ and $N_Y^{out}$ are the variances
of the total equivalent noise in the teleportation process, and unity transfer gain is
always assumed. If the state to be teleported is a minimum uncertainty state
(\eg coherent or squeezed state), our condition is identical to the Ralph and Lam
criterion : $T_X^{out} + T_Y^{out} > 1$, where $T_X^{out}$ and $T_Y^{out}$ 
are the SNR transfer coefficients.
For teleporting coherent states, the optimum efficiency is obtained when $N_X^{out}=N_Y^{out}$, 
and the fidelity must satisfy $F>2/3$. Though the result $F_{exp} = 0.58$ reported
in ref. \cite{FLB98} falls below that value, this experiment is nevertheless a very significant
achievement in defining and using the concept of continuous variables quantum teleportation. 

\section*{Acknowledgements}
This work was carried out in the framework of the european IST/FET/QIPC project ``QUICOV".
Useful discussions with Tim Ralph, Ping Koy Lam and Christine Silberhorn are acknowledged.

\begin{figure}
\vspace{10cm}
%\special{picture fig1tc}
\vspace{1cm}
\caption{Criteria for teleporting a coherent state using
a shared EPR state with limited squeezing and transmission efficiency $\eta$.
While the fidelity value $F_{g_T=1}$ is almost always larger than 0.5,
the product of equivalent input noises
($N_X^{out} N_X^{out} < 1$) or the information 
transfer efficiency ($T_X^{out} + T_Y^{out} > 1$) set more stringent limits
which can be fulfilled only when $\eta > 0.5$, with more than 3 dB squeezing
(thick part of the full lines). }
\end{figure}


\begin{references}

\bibitem{BBC93}
C.H. Bennett, G. Brassard, C. Crepeau, R.Jozsa, A. Peres, and W.K. Wootters, 
Phys. Rev. Lett. {\bf 70}, 1895 (1993).

\bibitem{BPM}  
D. Bouwmeester, J. W.
Pan, K. Mattle, M. Eibl, H. Weinfurter, and A. Zeilinger, Nature {\bf
390}, 575 (1997).

\bibitem{BBM}
D. Boschi, S. Branca, F. De Martini, L. Hardy, and S. Popescu,
Phys. Rev. Lett. {\bf 80}, 1121 (1998).

\bibitem{FLB98}
A. Furusawa, J.L. Sorensen, S.L. Braunstein, C.A. Fuchs, H.J. Kimble, and 
E.S. Polzik, Science {\bf 282}, 706 (1998).

\bibitem{BK98}
S.L. Braunstein and H.J. Kimble, Phys. Rev. Lett. {\bf 80}, 869 (1998).

\bibitem{RL98} 
T.~C.~Ralph and P.~K.~Lam, Phys. Rev. Lett.  {\bf 81}, 5668 (1998).

\bibitem{BK98b}
S.L. Braunstein and H.J. Kimble, Nature {\bf 394}, 840 (1998)

\bibitem{HCW90} M.~J.~Holland, M.~J.~Collett, D.~F.~Walls and
M.~D.~Levenson, Phys. Rev. A {\bf 42}, 2995 (1990).

\bibitem{PRG94} J.~-Ph.~Poizat, J.~-F.~Roch and P.~Grangier,
Ann. Phys. Fr. {\bf 19}, 265 (1994).

\bibitem{GLP} Ph. Grangier, J.-A. Levenson and J.-Ph. Poizat, 
	Nature {\bf 396}, 537 (1998)

\bibitem{Pr} John Preskill, Lecture Notes on Quantum Information
and Computation, Caltech (1998). 

\bibitem{epr} Z.Y. Ou, S.F. Pereira, H.J. Kimble, K.C. Peng,
Phys. Rev. Lett. {\bf 68}, 3663 (1992); Z.Y. Ou, S.F. Pereira, H.J. Kimble, 
Appl. Phys. {\bf 55}, 265 (1992) and references therein. 

\bibitem{rem} It can be shown that unity gain also appears as a ``best compromise"
between the two criteria of ref. \cite{RL98}.

\bibitem{pp}
L. Hardy, ``Disentangling Nonlocality and Teleportation",
arXiv:quant-ph/9906123.

M. Zukowski, ``Bell Theorem for Nonclassical Part of Quantum Teleportation
Process", arXiv:quant-ph/9912029.

L. Hardy argues that teleportation has nothing to do with non-locality,
and all to do with no-cloning. Since HI are essential for no-cloning,
our approach may be relevant from that point of view also.


\end{references}
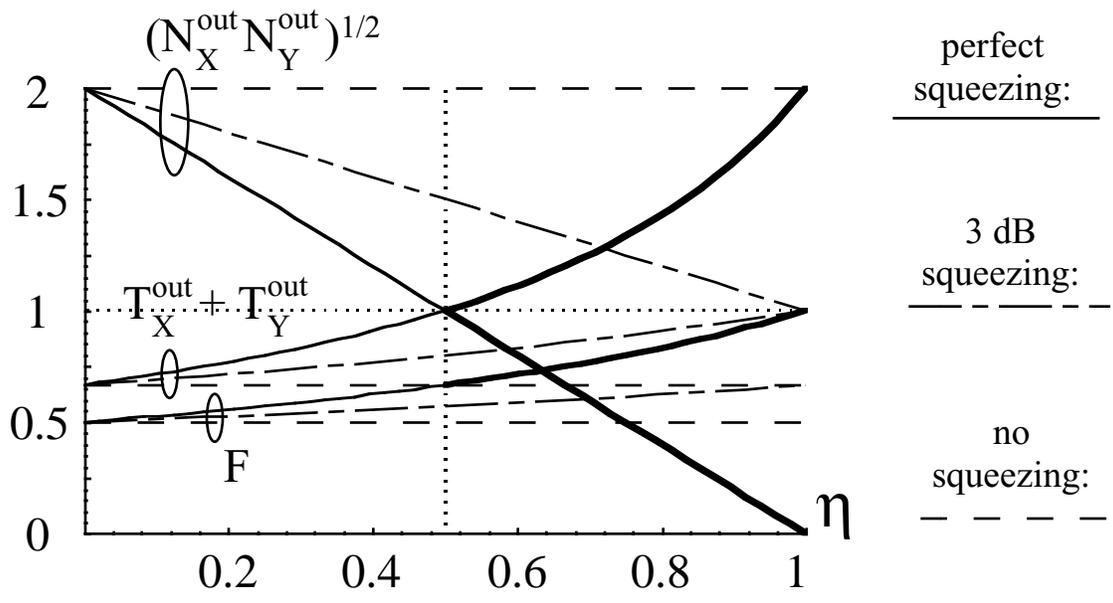
\end{document}